\documentclass[journal]{IEEEtran}
\usepackage{graphicx}
\usepackage{rotating}
\usepackage{longtable}
\usepackage{multirow}
\usepackage{lscape}
\usepackage{pifont}
\usepackage{rotating}
\usepackage{tablefootnote}
\usepackage{tabularx}
\usepackage[T1]{fontenc}
\usepackage[utf8]{inputenc}
\usepackage{csquotes} 
\usepackage[font=footnotesize,labelfont=bf]{caption}
\usepackage[font=footnotesize,labelfont=bf]{subcaption}
\usepackage{graphicx}
\usepackage{adjustbox}
\usepackage{amsmath,amssymb,amsfonts}
\usepackage{algorithm,algorithmicx,algpseudocode}
\algrenewcommand\algorithmicrequire{\textbf{Input:}}
\algrenewcommand\algorithmicensure{\textbf{Output:}}
\algdef{SE}[DOWHILE]{Do}{doWhile}{\algorithmicdo}[1]{\algorithmicwhile\ #1}%
\usepackage{url}

\usepackage{hyperref}
\usepackage{array}
\usepackage{color, colortbl}
\definecolor{ashgrey}{rgb}{0.7, 0.75, 0.71}
\definecolor{violetblue}{rgb}{0.2,0.29,0.7}
\definecolor{mediumvioletred}{rgb}{0.78,0.08,0.52}

\begin{document}
%


\title{On the Role of Hash-based Signatures in Quantum-Safe Internet of Things: Current Solutions and Future Directions}

%
%


\author{Sabah~Suhail, Rasheed~Hussain, Abid~Khan, and~Choong~Seon~Hong,~\IEEEmembership{Senior~Member,~IEEE}
\thanks{S. Suhail and C. S. Hong are with Department of Computer Science and Engineering, Kyung Hee University, South Korea (e-mail:{sabah,cshong}@khu.ac.kr).}
\thanks{R. Hussain is with Networks and Blockchain Lab, Innopolis University, Russia (e-mail:{r.hussain@innopolis.ru}).} 
\thanks{A. Khan is with Department of Computer Science, Aberystwyth University, United Kingdom (email: abk15@aber.ac.uk).}}

\maketitle

\begin{abstract}
The Internet of Things (IoT) is gaining ground as a pervasive presence around us by enabling miniaturized \emph{\enquote{things}} with computation and communication capabilities to collect, process, analyze, and interpret information. Consequently, \emph{trustworthy data} act as fuel for applications that rely on the data generated by these things, for critical decision-making processes, data debugging, risk assessment, forensic analysis, and performance tuning. Currently, secure and reliable data communication in IoT is based on public-key cryptosystems such as Elliptic Curve Cryptosystem (ECC). Nevertheless, reliance on the security of de-facto cryptographic primitives is at risk of being broken by the impending quantum computers. Therefore, the transition from classical primitives to quantum-safe primitives is indispensable to ensure the overall security of data en route. 
In this paper, we investigate applications of one of the post-quantum signatures called \emph{Hash-Based Signature} (HBS) schemes for the security of IoT devices in the quantum era. 
We give a succinct overview of the evolution of HBS schemes with emphasis on their construction parameters and associated strengths and weaknesses. Then, we outline the striking features of HBS schemes and their significance for the IoT security in the quantum era. We investigate the optimal selection of HBS in the IoT networks with respect to their performance-constrained requirements, resource-constrained nature, and design optimization objectives. In addition to ongoing standardization efforts, we also highlight current and future research and deployment challenges along with possible solutions. Finally, we outline the essential measures and recommendations that must be adopted by the IoT ecosystem while preparing for the quantum world.
\end{abstract}

\begin{IEEEkeywords}

Blockchain, Hash-based signature, Internet of Things, Public-key cryptography, Quantum computing.

\end{IEEEkeywords}

\IEEEpeerreviewmaketitle

\section{Introduction} \label{introduction}
\IEEEPARstart{T}{he} proliferation of cost-effective miniaturized devices with computation and communication capabilities is providing promising solutions to enhance the quality of life and style in a plethora of ubiquitous application areas including, but not limited to, smart cities, meteorology, health-care systems, smart grid, industrial automation, and precision agriculture. These devices with the afore-mentioned capabilities, together constitute the Internet of Things (IoT)~\cite{al2015internet}.
Regardless of such comforts, the revolutionary IoT technology is vulnerable to security glitches that arise due to the interconnection of unattended and globally accessible things with the untrusted and unreliable Internet. Loopholes in the system infrastructure lure adversaries to launch different attacks; for example, data forging, Sybil attacks, false data injection, replay attacks, and denial of participation. Such attacks will have catastrophic consequences for the high-assurance applications that are involved in crucial decision-making processes based on aggregated sensor data (such as health-care, industrial, and financial applications)~\cite{Hassija2019, Neshenko2019}. Thus, to provide data authenticity and protection against data forgery, potential countermeasures for IoT security are essential elements for ensuring authentic and trustworthy data acquisition and data communication.

\begin{table*}[ht!]
  \centering
\caption{Acronyms and their explanation.} \label{tab:notations}
  \begin{tabular}{|p{2.0cm}|p{5cm}|p{2.0cm}|p{5cm}| }
  \hline
\textbf{Acronym} & \textbf{Explanation} & \textbf{Acronym} & \textbf{Explanation}   \\ [0.5ex]
 \hline 
 \hline
IoT & Internet of Things & HBS & Hash-Based Signature   \\
 \hline
OTS & One-time Signature & WOTS & Winternitz OTS \\
 \hline
WOTS\textsuperscript{PRF} & WOTS (Pseudo Random Function) & MTS & Multi-time Signature\\
 \hline
MSS & Merkle Signature Scheme & XMSS & Extended MSS\\
 \hline
HS & Hierarchical Signature & LMS & Leighton Micali Scheme \\
 \hline
XMSS-T & XMSS with tightened security & XMSS\textsuperscript{MT}& XMSS (Multi Tree)\\
 \hline
FTS & Few-Time Signature & HORS & Hash to Obtain Random Subset \\ 
 \hline
PORS & PRNG to Obtain Random Subset & HORS-T & HORS (with Tree) \\
 \hline
DLT & Distributed Ledger Technology & IIoT & Industrial IoT \\
 \hline
PQC & Post Quantum Cryptography  & PRNG & Pseudo-Random Number Generator  \\
 \hline
 QRNG & Quantum Random Number Generation & QKD & Quantum Key Distribution \\
 \hline
\end{tabular}
\end{table*}

Security protocols usually rely on the cornerstone applications of digital signatures for \emph{authentication}, \emph{integrity}, and \emph{non-repudiation}. For instance, code signing of devices for software and firmware to ensure legitimate updates or upgrades in software suites or patches, Distributed Ledger Technology (DLT) to ensure valid cryptocurrency transactions, Vehicular Ad hoc NETwork (VANET) to ensure trustworthy message communication among vehicles or road-side units, and medical implantable and wearable sensors for data integrity, use digital signatures.

In these real-world scenarios, the most widely used cryptographic schemes for digital signatures are RSA \cite{rivest1978method}, Digital Signature Algorithm (DSA)~\cite{elgamal1985public}, and Elliptic Curve Digital Signature Algorithm (ECDSA)~\cite{johnson2001elliptic}. Security of these classical cryptographic primitives relies on the hardness of factoring integers and computing discrete logarithms~\cite{cheng2017securing}. However, it is expected that with the not-so-far arrival of quantum computers, these computational problems will be susceptible to quantum computer cryptanalysis using Shor’s quantum algorithm~\cite{shor1999polynomial} 
and variational quantum factoring~\cite{anschuetz2019variational} and therefore, can be solved by quantum computers in polynomial time. Doubling the key length increases the difficulty; however, this is not enough for a sustainable edge. Furthermore, Grover's algorithm~\cite{grover1996fast} can allow brute-force attacks to address the effect of quantum computing on symmetric cryptography.

In the interim, security mechanisms of digital signatures not only coerce the need for rigorous scrutiny to thwart both classical and post-quantum attacks but also call for state-of-the-art security solutions for resource-constrained and performance-constrained IoT devices to continue utilizing the IoT-based services in the quantum world.
Therefore, the inexorable march of quantum hype entails dependable quantum-safe digital signature schemes. In this regard, Hash-Based Signature (HBS) schemes~\cite{mcgrew2016state} are promising candidates, offering security proofs relative to plausible properties of the hash, and the object of leading-edge standardization efforts.

\begin{table*}[ht!]
  \centering
\caption{Existing surveys and articles.} \label{tab:survey}
  \begin{tabular}{|p{0.8cm}|p{1.0cm}|p{4.0cm}|p{3.5cm}|p{5.0cm}| }
  \hline

\textbf{Year} & \textbf{Paper} & \textbf{Topic(s) of the article/survey} & \textbf{Related content in our paper} & \textbf{Enhancements in our paper}    \\ [0.5ex]
 \hline
\hline
2015 & \cite{butin2015real} & 
Further advantages of hash-based signatures, Obstacles to Widespread Use, Bridging the Gap & Section~\ref{standard}, Section~\ref{technical} & Coverage of technical, non-technical, and social challenges along with possible solutions to the respective problems in case of both stateful and stateless HBS schemes from IoT design and implementation perspective; Current state-of-the-art standardization efforts and industrial scale implementation efforts.  \\
 \hline
2016 & \cite{mcgrew2016state} & 
Stateful Hash-Based Signature Schemes
One-time, State Synchronization Security Risks, Overhead for hash-based signatures & Section~\ref{schemes}, Section~\ref{technical} & Overview of stateful and stateless HBS along with detailed taxonomy; In-depth discussion on technical and non-technical challenges particularly in the context of IoT.\\
 \hline
2017 & \cite{butin2017hash} & 
Hash-Based Signature Basics, Challenges and trade-offs & Section~\ref{schemes}, Section~\ref{features}, Section~\ref{standard}, Section~\ref{technical} & Overview of stateful and stateless HBS along with detailed taxonomy; Up-to-date standardization efforts; Coverage of HBS features from the perspective of IoT domain.\\
 \hline
2017 & \cite{cheng2017securing} &
Ongoing projects and developments
& Section~\ref{standard} & Up-to-date standardization efforts including the state-of-the-art industrial-scale efforts.  \\
 \hline
2018 &  \cite{palmieri2018hash} & 
Hash-based signatures, Challenges
& Section~\ref{standard}, Section~\ref{technical} & Up-to-date standardization efforts; Detailed technical and non-technical challenges.  \\
\hline
\end{tabular}
\end{table*}

\subsection{Existing Literature}
To date, not many surveys have been conducted that investigate various aspects of Post Quantum Cryptography (PQC). To the best of our knowledge, most of the existing surveys and articles focus on various sparse aspects of post-quantum digital signature schemes such as providing only a panoramic view of schemes, covering only technical details without connecting them with any application domain, schemes (excluding HBS) in combination with IoT, presenting HBS signatures basics without further exploring their association to any domain, and hence are indirectly related to the HBS-driven IoT. By narrowing down our survey to HBS schemes featuring IoT applications and focusing on more high-level issues, we present a holistic approach towards HBS in combination with IoT. 

Starting with the most relevant paper, in~\cite{palmieri2018hash}, the authors investigate the role of HBS schemes with a focus on underlying challenges in the IoT domain. Similarly, in~\cite{butin2017hash}, the authors provide an overview of post-quantum signature schemes with an emphasis on the basic structure of HBS schemes along with a few example schemes from each category (i.e., stateless and stateful), features, and standardization. \cite{butin2015real} lays out the obstacles to the widespread use of HBS in general. Besides, the authors discuss the efforts needed by the cryptographic research community to focus on the significance of standardization and integration in commonly used cryptographic software libraries and security protocols to support the broad adoption of HBS in the real world.
On the other hand, some works mostly cover the technical aspects (optimizing schemes through construction parameters, mathematical analysis, and performance evaluation through implementation on IoT platforms) of HBS schemes. These works include~\cite{mcgrew2016state} (discusses the problem of state management and provides possible solutions to solve it), \cite{aumasson2018improving} (discusses optimization of stateless HBS schemes), and \cite{pereira2016shorter, rohde2008fast, ghoshlightweight} (implement and evaluate proposed schemes on IoT devices), to name a few. 
Lastly, the focal point of the existing articles includes other classes of post-quantum signature schemes in the view of the IoT domain. For example,~\cite{lohachab2020comprehensive} discuss the role of PQC in IoT and associated open challenges. \cite{cheng2017securing} focuses on lattice-based and multivariate polynomial-based algorithms for constrained devices and networks. Similarly,~\cite{liu2018securing} emphasize the suitability of lattice-based cryptography by securing the communication between IoT and edge devices. Table~\ref{tab:survey} presents a summary of these surveys and their differences with our survey.

\subsection{Scope of This Survey}
In this paper, we present a comprehensive and systematic review of state-of-the-art technical, non-technical, and social issues that arise due to the integration of IoT in HBS schemes. The main contributions of our paper are summarized as follows.
\begin{itemize}
    \item Starting from the potential grounds for the transition to post-quantum signature schemes, we discuss the key questions to elaborate the reasoning behind this transition and further actions. Then, we provide a high-level working of the family of HBS schemes categorized as stateless, stateful, and hybrid based on key generation, signature generation, and other construction parameters. Along with the evolution of HBS schemes, we also highlight the strength and weaknesses of the respective schemes.  
    \item We focus on the features of HBS schemes and their significance for securing the application-dependent and platform-dependent IoT.
    \item With reference to IoT-driven use-cases, we present various elemental factors that must be considered while introducing HBS schemes in the IoT ecosystem. 
    \item In addition to the on-going standardization efforts and state-of-the-art industrial efforts, we provide an in-depth review of various research challenges such as technical, non-technical, and social challenges. We also map such requirements from IoT perspectives, highlight the open-ended challenges that need to be addressed by the research community, and finally outline recommendations to prepare and act strategically while moving towards the quantum era. 
\end{itemize}

The rest of the paper is organized as follows: Table~\ref{tab:notations} lists all the acronyms used in the paper. Section~\ref{overview} covers HBS schemes by including a quick high-level overview of the different types of stateful and stateless HBS schemes. The peculiar features of HBS schemes and their significance for the IoT domain are outlined in Section~\ref{features}. Considering the constraints of IoT devices, the usage of HBS in the IoT environment is presented in Section~\ref{HBSSIOT}. Section~\ref{challenges} describes the technical, non-technical, and social challenges and requirements of HBS schemes. Finally, Section~\ref{conclusion} concludes the paper.

\begin{figure*}[ht!]
\centerline{\includegraphics[width=4.5in]{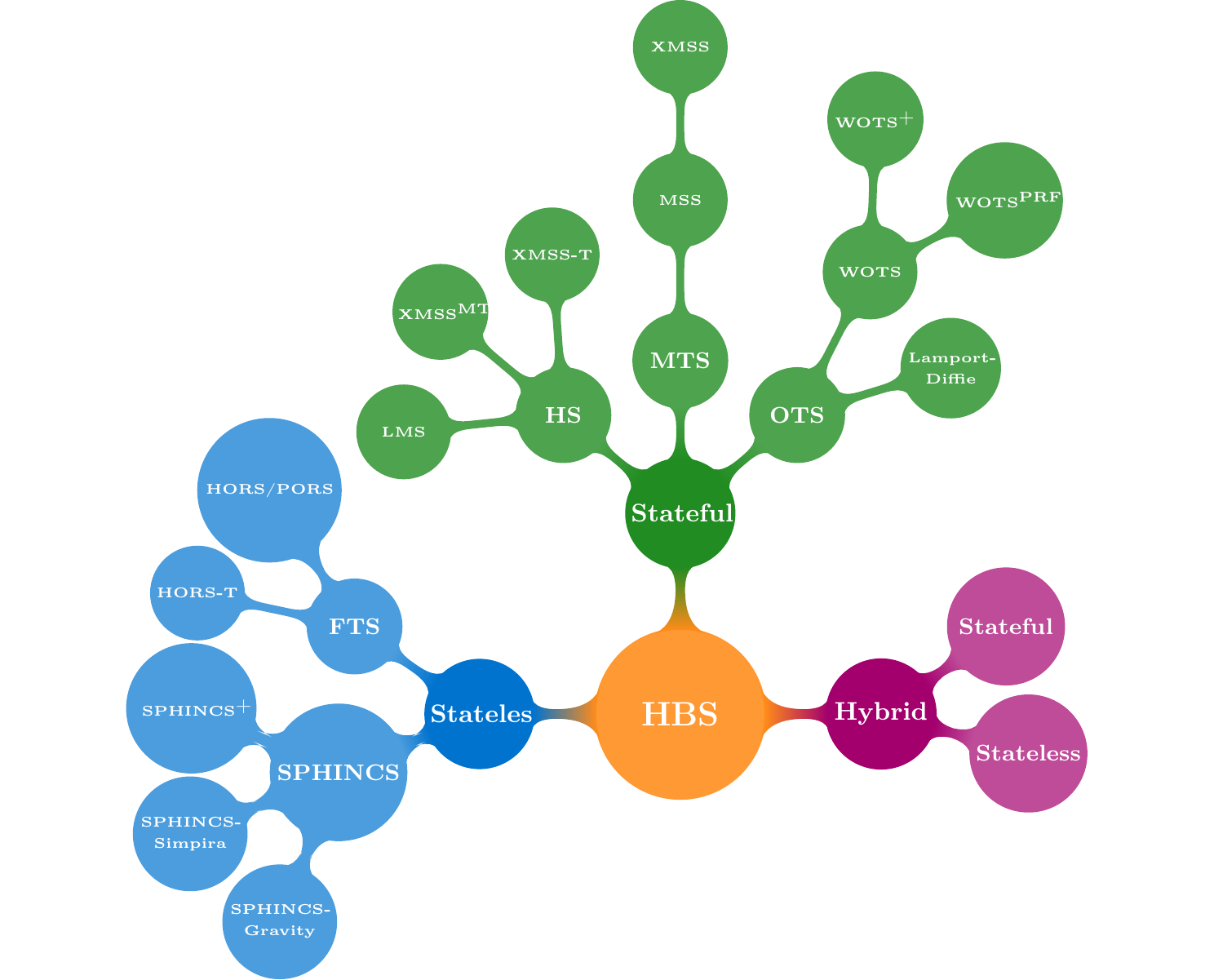}}
\caption{Taxonomy of hash-based signature schemes.} \label{mindmap}
\end{figure*}

\section{Transition from Traditional Digital Signatures to Hash-Based Signatures}
\label{overview}
Starting with the limitations of traditional digital signature schemes due to looming threats by quantum computing to traditional cryptographic solutions, in this section, we present the potential reasons for the transition to quantum-secure schemes. Then, we discuss quantum-safe security solutions as HBS schemes. We provide a quick overview of the evolution of HBS schemes to address the problems of key generation, signature generation, signature verification, etc.

\begin{table*}[!ht]
\centering
\caption{Examples of widely deployed cryptographic systems for 128-bit pre-quantum security level.} \label{tab:PK_algo}
\resizebox{\textwidth}{!}{%
 
\begin{tabular}{|l|l|l|l|l|l|l|l|l|}
\hline
\multicolumn{1}{|c|}{\textbf{Class}}                                                                  & \multicolumn{6}{c|}{\textbf{Public-key Cryptography}}                                                           & \multicolumn{2}{l|}{\textbf{Symmetric Cryptography}} \\ \hline \hline
\textbf{Cryptographic primitives}                                                                     & Integer factorization & \multicolumn{2}{l|}{Discrete logarithm} & \multicolumn{3}{l|}{Elliptic curves} & \multicolumn{2}{l|}{}                       \\ \hline 
\textbf{Cryptosystems}                                                                                & RSA                   & DH                 & DSA                & Elgamal      & ECDH      & ECDSA     & AES       & SHA-256 (pre-image security)    \\ \hline 
\textbf{\begin{tabular}[c]{@{}l@{}}Post-quantum security level \\ (broken by algorithm)\end{tabular}} & Shor                  & Shor               & Shor               & Shor         & Shor      & Shor      & Grover    & Grover                          \\ \hline 
\end{tabular}
}%
\end{table*}

\subsection{Limitations of Classical Digital Signature Schemes} 
The end of traditional cryptosystems is marked by the Shor's and Grover's algorithms. On one hand, the Shor's algorithm solves the underlying mathematical problems of public-key algorithms (as mentioned in Table~\ref{tab:PK_algo}) whereas, on the other hand, Grover’s algorithm can reduce the effective security strength of algorithms (such as the Advanced Encryption Standard (AES)~\cite{standard2001announcing} and 3-DES (Triple Data Encryption Standard)~\cite{barker2017recommendation}) to roughly half for a given key length, thereby rendering infrastructures secured by them vulnerable to exploitation~\cite{mulholland2017day}. 

With the proliferation of quantum computing technologies, the epoch-making incident of the end of the currently used classical digital signature scheme in the foreseeable future raises the following concerning questions. 
The first question is that despite conjectured security of the underlying cryptographic mechanisms, \emph{why the traditional signature schemes are unable to withstand the quantum computers?} Crudely put, the exponential speed-up brought about by quantum computer stems from the fact that it acts as a massively parallel computer which is made possible by quantum mechanics called \emph{superposition} (i.e., the ability for a quantum bit (qubit) to be both a one and a zero at the same time). Thus, proper implementation of superposition state in a quantum computer can provide exponential computing power which may break all existing schemes.

The second question is, \emph{what will happen if all the current cryptographic security solutions suddenly become ineffective?} The failure of classical cryptosystems may have a devastating effect on the systems and may lead to the destruction of the security fabric that connects much of the omnipresent IoT world today and in the near future. Thus, in addition to other domains, the IoT applications that rely on pivotal features of existing digital signatures, principles of data integrity, message authentication, and non-repudiation, are going to have profound aftermath on sensory data in terms of security and privacy.  

The third question is \emph{when such a dilemma is going to happen?}
According to the experts at the University of Waterloo, there is a 1-in-7 chance of these cryptographic primitives being affected by quantum attacks in 2026, and a 1-in-2 chance by 2031~\cite{chance2031}.

Finally, the fourth question is, \emph{what to do now?} To provide security to IoT applications, quantum-safe schemes are explored by academia and industry. The post-quantum signature schemes can be classified into five categories as i) Hash-based ii) Lattice-based iii) Multivariate polynomial based iv) Code-based, and v) Super-singular isogeny based schemes. Among these quantum-secure signature schemes, we opted for HBS schemes because they are well-studied schemes with minimal security requirements, practiced, reasonably fast, yield small size signatures, and have strong security guarantees, to name a few. 
The afore-mentioned discussion calls for the transition to quantum-secure algorithms to ensure adequate cryptographic protections in the hyper-connected IoT world. In the following, we dive a bit deeper into the stateless and stateful HBS schemes.

\subsection{HBS Schemes: From Stateful to Stateless} 
\label{schemes}
The design principle of HBS is to leverage an underlying cryptographic secure hash function that exhibits any of the security property including \emph{one-wayness}, \emph{pre-image resistance}, \emph{second-preimage resistance}, and \emph{collusion resistance}.
Based on the implementation approach, HBS schemes can be classified as \textit{stateless} and \textit{stateful} schemes which can be further categorized as \textit{One-Time Signature} (OTS), \textit{Few-Time Signature} (FTS), \textit{Multi-Time Signature} (MTS), and \textit{Hierarchical Signature} (HS), depending on key generation, signature generation, and other construction parameters. Fig.~\ref{mindmap} represents the detailed classification of stateful and stateless HBS schemes. In the following, we further elaborate on these categories.

\subsection{Stateful HBS Schemes}
A stateful digital signature scheme necessitates the maintenance of the updated non-repeated secret key upon each signature generation process. It is essential to keep track of non-repeated key pairs, failing which will result in the degradation of the security of the cryptographic scheme. Different categories of stateful schemes are given as follows:

\subsubsection{Stateful One-time Signature Schemes (OTS)}
Among the stateful signature schemes, OTS schemes form a fundamental building block for HBS. Common examples of seminal OTS are Lamport-Diffie scheme~\cite{lamport1979constructing}, Winternitz scheme~\cite{dods2005hash}, and its variants WOTS\textsuperscript{+}~\cite{hulsing2013w}, WOTS\textsuperscript{PRF}. To sign a message with OTS schemes, the private key is uniformly generated at random, whereas the public key is derived as a function of the private key, involving the underlying hash function. 

Lamport-Diffie scheme provides very strong security on minimal assumptions; however, it has some major downsides
which prevented its wide adoption. Firstly, it is one-time, making it in-apposite for the majority of use cases of digital signatures. Secondly, the keys and the signatures are extremely large (as shown in Table~\ref{tab:OTS/FTS}).
The deterring issue of extremely large key length and signature size in the Lamport-Diffie scheme is resolved through WOTS by introducing a Winternitz parameter that controls time/memory trade-off. Therefore, reducing the space required for keys and signatures makes WOTS a good choice for memory-constrained embedded devices (and hence IoT), but at the cost of slower signing and verifying process.  
Overall, OTS schemes are single-use in nature (i.e., can only sign a pre-defined number of messages with a key pair, which introduce a key renewal overhead) and therefore inadequate to use in real-world applications. This is because using the same key multiple times may enable an attacker to reveal more parts of the private key, and hence compromise the security of the underlying scheme.

\begin{figure}[ht!]
\centerline{\includegraphics[width=3.5in]{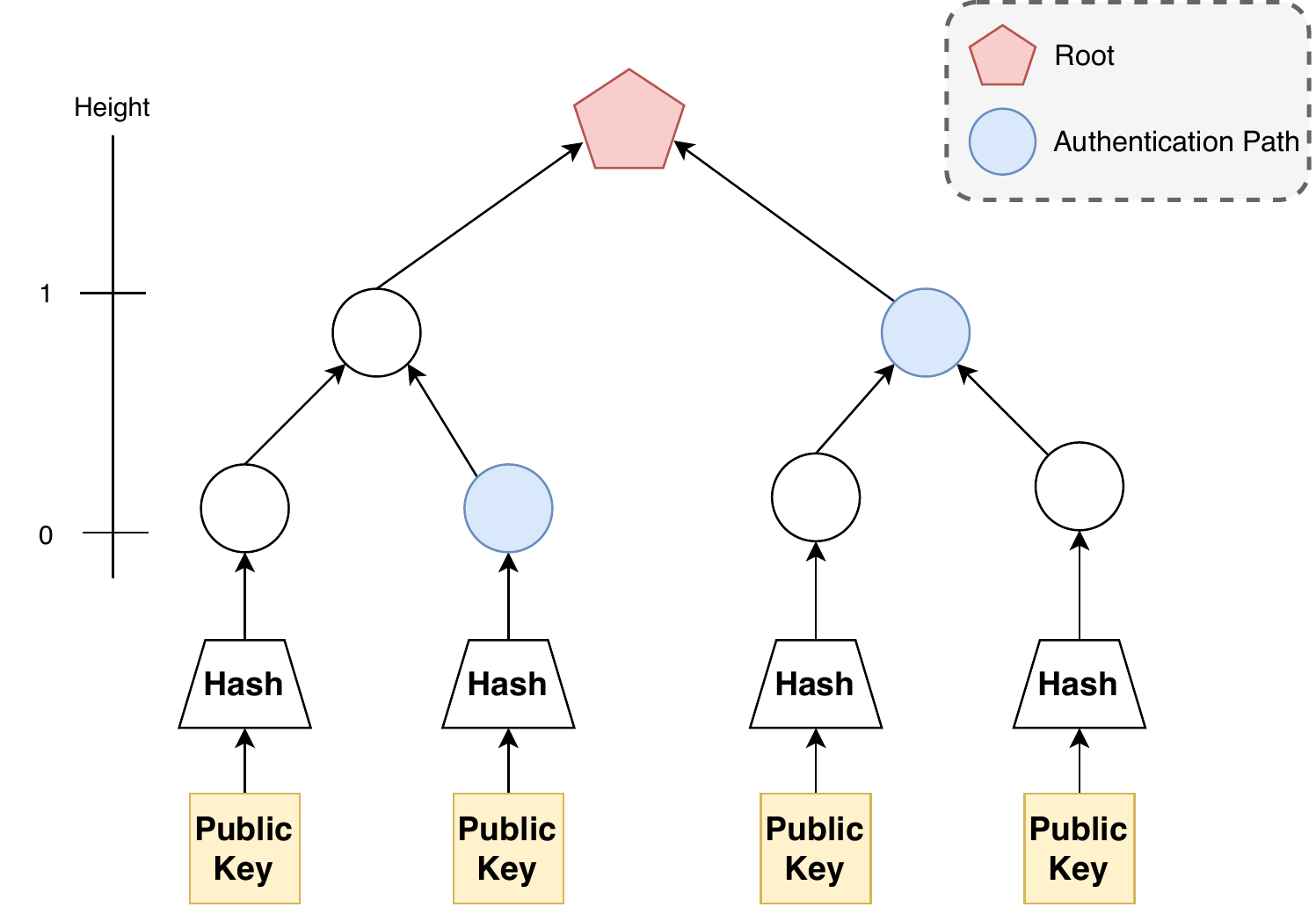}}
\caption{\emph{Merkle Signature Scheme} (MSS) using One-Time Signature (OTS): An illustration of stateful Multi-time Signature (MS) scheme. (Figure adapted from \cite{hulsing2013practical}.)}
\label{fig:mss}
\end{figure}

\begin{figure*}[!ht]
\centerline{\includegraphics[width=4.0in]{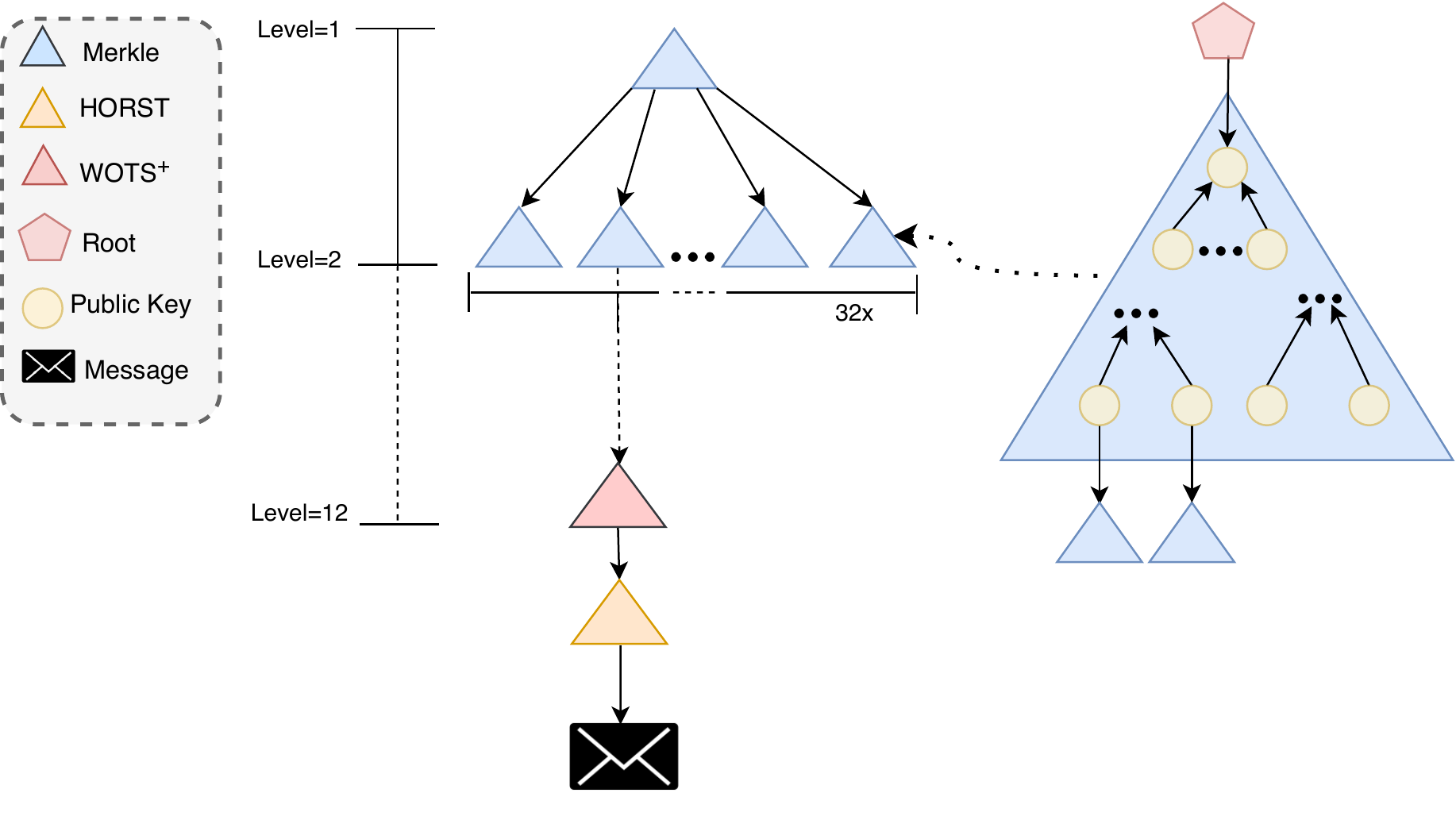}}
\caption{Hypertree structure used in \emph{SPHINCS}: An illustration of stateless Hierarchical Signature (HS) scheme. (Figure adapted from~\cite{kolbl2018putting}).}
\label{fig:sphincs}
\end{figure*}   

\subsubsection{Stateful Multi-time Signature Schemes (MTS)}
To untangle the peculiarity of the one-time nature of OTS schemes, MTS schemes are proposed to construct many-time signatures by using OTS as an under-structure. In~\cite{merkle1989certified}, Ralph Merkle proposed Merkle Signature Scheme (MSS) to generate multiple aggregated public and private keys by combining a large number of OTS key pairs into a single binary hash tree structure (as shown in Fig.~\ref{fig:mss}). To authenticate the relation of a one-time public key with the global public key (also referred to as tree root), signatures keep on appending a sequence of intermediate tree nodes, called authentication paths (as shown in Fig.~\ref{fig:mss}). Such paths allow the validator to reconstruct the path from the relevant one-time public key to the tree’s root upon signature verification. To enhance the efficacy and practicability of MSS, the following optimization strategies are adopted based on different flavors of Merkle tree construction, leaves calculation, and parameter specifications.
Firstly, the global private key can be efficiently constructed by using a cryptographically secure Pseudo-Random Number Generator (PRNG) such that from an initial seed value (which acts as a private key),
both successive seeds and one-time secret keys are derived. Thus, in lieu of storing all OTS secret keys, it is sufficient to store only the seed value of the PRNG, while generating other seed values on-the-fly. It ultimately minimizes storage requirements. Such a strategy for global private key construction also provides forward secrecy and existential unforgeability under adaptive chosen message attack~\cite{kannwischer2017physical}. Nevertheless, it necessitates precise counter management for tracking the used keys, particularly across multiple invocations of signing algorithm, because using any one-time private key twice is imperative to security. Secondly, the performance-optimized BDS algorithm~\cite{buchmann2009hash} is used for efficient computation of authentication path such that it caches the authentication path from the previous signature, thus, instigate time/memory trade-off. To this end, concrete examples of M-time signature schemes are Extended Merkle Signature Scheme (XMSS)~\cite{buchmann2011xmss}, ~\cite{leighton1995large}, and ~\cite{buchmann2007merkle}.

\subsubsection{Stateful Hierarchical Signature Schemes (HS)}
Although the use of the optimized BDS algorithm provides sufficient performance during the signature generation of XMSS implementation, it is still relatively slower in generating a new key pair due to the requirement of constructing the entire hash tree~\cite{kannwischer2017physical}. Hence, to further improve performance, HS schemes are proposed. In essence, HS schemes are MTS schemes that use other hash-based signatures in its construction. The idea of HS is based on the formation of a hyper-tree that involves tree chaining by using multiple layers of MSS tree. In this form of Merkle tree construction, the upper layers are used to sign the roots of the layers below while only the lowest layer is used to sign messages. 
Notable examples of HS are XMSS-Multi Tree (XMSS\textsuperscript{MT})~\cite{hulsing2013optimal}, XMSS with tightened security (XMSS-T)~\cite{hulsing2016mitigating}, and Leighton Micali Scheme (LMS)~\cite{leighton1995large}. XMSS\textsuperscript{MT} is particularly ideal for applications that require virtually a large number of messages to be signed. 
Note that, XMSS\textsuperscript{MT} should be used in conjunction with other optimization strategies, including the BDS algorithm, PRNG, and caching of the authentication paths, otherwise the required storage and the long time for random number generation outweigh the performance gain of XMSS\textsuperscript{MT}.
Additionally, the LMS has two variants, i.e., Leighton Micali one-time signature (LM-OTS) and the many-time signature scheme LMS~\cite{mcgrew2016hash}.

\subsection{Stateless HBS Schemes}
Keeping track of the last used OTS key pair is considered to be one of the major downsides of stateful schemes. To address this intriguing problem, stateless schemes are introduced. A stateless digital signature scheme eliminates the need for maintaining the updated non-repeated secret key upon each signature generation process. Because unlike OTS schemes (WOTS or its variants), stateless HBS schemes use few-time signature schemes, for instance, Hash to Obtain Random Subset/PRNG to Obtain Random Subset (HORS/PORS)~\cite{reyzin2002better} and HORS with Tree (HORS-T)~\cite{bernstein2015sphincs}.

\subsubsection{Stateless Hierarchical Signature Schemes (HS)}
Some of the examples of the stateless HS scheme are SPHINCS~\cite{bernstein2015sphincs} and its variants SPHINCS-Simpira~\cite{gueron2017sphincs}, Gravity-SPHINCS~\cite{aumasson2018improving}, and SPHINCS\textsuperscript{+}~\cite{bernstein2019sphincs+}.
Similar to XMSS\textsuperscript{MT}, SPHINCS uses a hypertree such that the upper layers use XMSS with WOTS\textsuperscript{+} to sign roots of their ancestors, while the lowest layer uses a Merkle tree construction with HORS-T for signing messages(as shown in Fig.~\ref{fig:sphincs}). Since the stateless schemes do not keep a record of used key pairs, hence to ensure the correct few-time usage of key pairs, SPHINCS deploys multiple HORS-T key pairs and selects a random one for each signature generation. As a result, no path-state tracking is required.

Generating all HORS-T and WOTS\textsuperscript{+} private keys with a PRNG for key generation and computing one tree in each layer for signature generation results in the feasible computation for SPHINCS. Nevertheless, stateless schemes pose the following performance issues. Firstly, the signature generation is more expensive because the key pairs are used in random order rather than successive order; hence, the optimization algorithm BDS is no longer suitable. Secondly, in contrast to WOTS\textsuperscript{+}, HORS-T signatures are relatively much larger.
We summarize the stateless and stateful class of HBS schemes along with their signature size, key length, and other relevant details in Table~\ref{tab:OTS/FTS} and Table~\ref{tab:schemes}.

\begin{table*}[!ht]

\caption{OTS/FTS schemes for 384-bit message length and 128-bit (approximately) post-quantum security level.} \label{tab:OTS/FTS}
  \begin{tabular}{|p{3.5cm}|p{3.5cm}|p{3.5cm}|p{3.5cm}| }
  \hline

\textbf{Scheme} & \textbf{Type} & \textbf{Signature size (KB)} & \textbf{Key size (KB)}  \\ [0.5ex]
 \hline 
 \hline
Lamport-Diffie & OTS & 18.4 & 36.9   \\
WOTS & OTS &  4.8  & 4.8   \\
WOTS\textsuperscript{PRF} & OTS  & 3.2  & 3.2   \\
WOTS\textsuperscript{+} & OTS  & 3.2  & 3.7   \\
\hline
HORS/PORS & FTS & 1.2  &  3.1 MB  \\
HORS-T & FTS & 17.3 & 0.05  \\
\hline
\end{tabular}
  
\end{table*}

\begin{table*}[!ht]

\caption{Stateful and Stateless hash-based signature schemes:a comparative summary}
\label{tab:schemes}

\small
\begin{tabular}{|p{1.7cm}|p{1.9cm}|p{1.3cm}|p{1.0cm}|p{2.4cm}|p{1.8cm}|p{1.8cm}|p{2.5cm}| }
  \hline

\textbf{Scheme} & \textbf{Instantiation} & \textbf{Message length} & \textbf{Type} &  \textbf{Base scheme} & \textbf{Key-reuse capacity} & \textbf{Signature size (KB)} & \textbf{Key size (KB)}  \\ [0.5ex]
 \hline
 \hline
MSS & SHA-384 & 384-bit & Stateful & WOTS & $2^{60}$ & 7.7  & 0.05  \\
XMSS & SHA-256 & 256-bit & Stateful & WOTS\textsuperscript{PRF} & $2^{60}$ & 4.7 & 0.03 \\
XMSS\textsuperscript{MT} & AES-128 & 256-bit & Stateful & WOTS\textsuperscript{PRF} & $2^{80}$ & 10.5 & private key = 26.1, public key = 1.8\\
 \hline
 
SHPINCS & SHA-256 & 512-bit & Stateless & HORS-T; WOTS\textsuperscript{+} & Unlimited & 41.0 & 1.0 \\
G-SPHINCS & Haraka 
& 512-bit & Stateless & PORS; WOTS & Unlimited & 30.0 &
private key = 0.06, public key = 0.03 \\ 
SPHINCS-S & Simpira 
& 512-bit & Stateless & HORS-T; WOTS\textsuperscript{+} & Unlimited & 41.0 & 1.0 \\
\hline
\end{tabular}
\end{table*}

\section{Features of HBS Schemes and Their Significance in The IoT Environment} 
\label{features}
\begin{figure}[ht!]
\centerline{\includegraphics[width=4.0in]{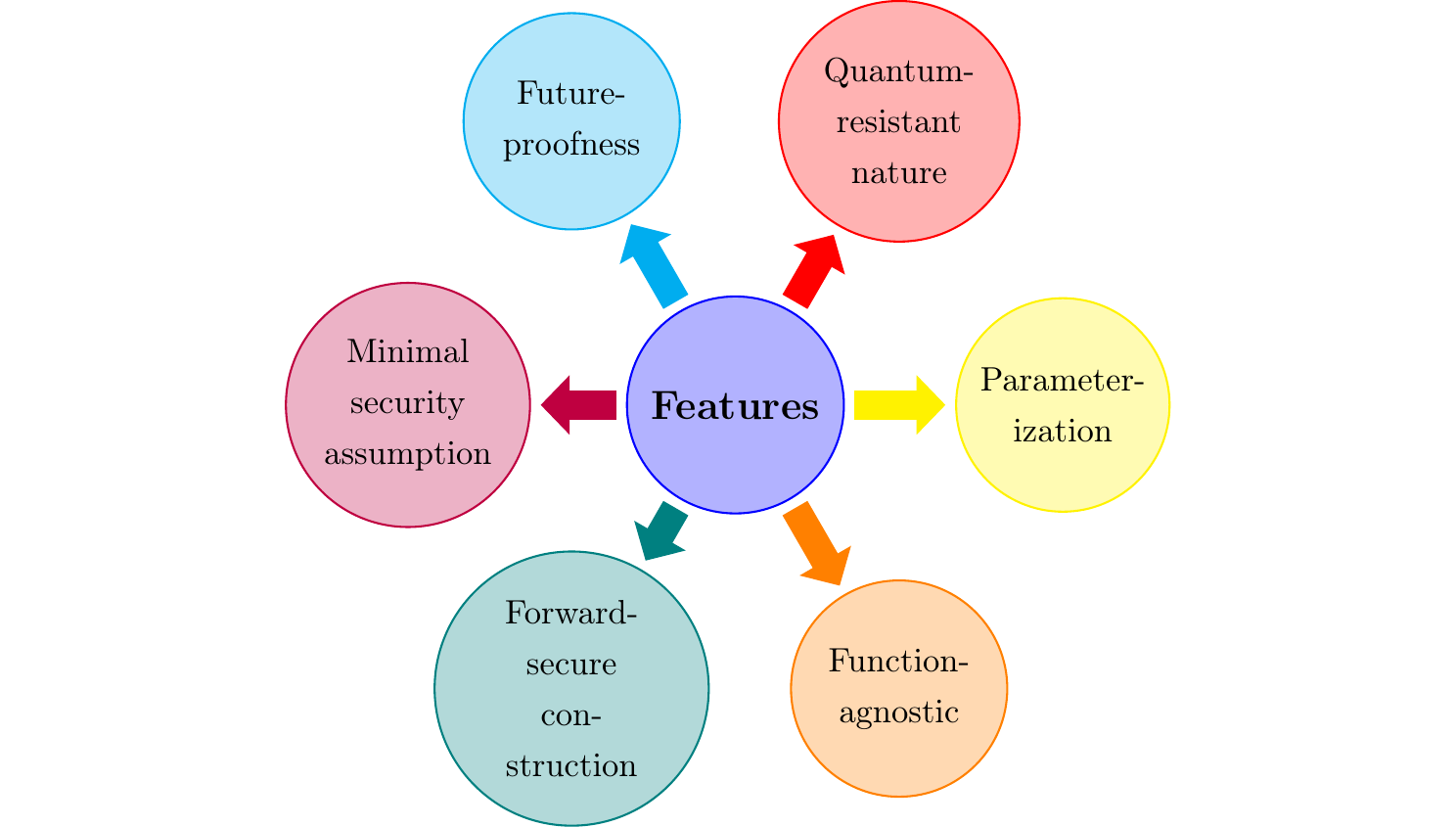}}
\caption{Striking features of hash-based signature schemes.}
\label{fig:features}
\end{figure}

Several arguments underpinning the use of HBS schemes in the IoT ecosystem include quantum-resistance, minimal security assumptions, function agnostic, forward-secure construction, and extensive tunable parameters. In this section, we elaborate on the features of HBS schemes by associating their felicitous illustrations for the IoT environment. The striking features of HBS schemes are summarized in Fig.~\ref{fig:features}.

Traditional signature schemes generally require consideration of number-theoretic hardness assumptions (such as composite integer factorization and discrete logarithm problem) in addition to the security of hash functions. On the contrary, HBS schemes solely rely on the underlying secure cryptographic hash function, thereby pruning the attack surface and reducing the opportunities for cryptanalysis. For instance, a secure implementation of XMSS exclusively depends on a secure cryptographic hash function that is either second preimage resistant or pseudorandom to be secure. Thus, the idea of \emph{minimal security assumption} in HBS effectively reduces the complexity of implementation by eliminating the reliance on multiple security components. Hence, it streamlines the deployment among diverse implementations (such as massively heterogeneous applications seem good candidates) and devices (such as resource-constrained IoT devices)~\cite{buchmann2011xmss}.

HBS schemes are \emph{function-agnostic}, i.e., they can be built on top of any hash function that satisfies the security requirements of cryptographic hash functions. Such inherent flexibility of HBS allows the selection of different underlying hash functions to meet the desired performance requirements depending on the application-specific environment. The function-agnostic and \emph{quantum-resistant} nature of HBS schemes make them \emph{future-proof} such that the underlying hash functions can be simply substituted (in terms of implementation) in case of vulnerabilities with any of the specific hash function over time. For instance, to handle a multi-target attack, the researchers shift to collision-resilient signature schemes as collision resistance is subject to birthday attacks in comparison to preimage and second-preimage resistance~\cite{halevi2006strengthening}. 

The feature of future-proofness manifests long-term security of lifetime devices. 
One aspect of such scenarios is the hardware protection of multitude field-deployed devices in massive IoT. For example, the deployment of new sensor motes in industrial automation, agriculture precision, environment monitoring, and other mission-critical applications are deleterious, costly, and time-consuming task; therefore, the hardware longevity must be considered to address future threats.
Another aspect is high assurances of digitally-signed firmware to prevent adversaries from stealing the signing credentials of long-running devices. 
Another example includes mission-critical devices that require data trustworthiness, especially for applications that perceive the value of sensor data for decision-making processes, risk assessment, and performance evaluation~\cite{suhail2019orchestrating}.
Under both aspects, long-term security offered by the PQC in the form of hash-based signatures must be adopted to ensure trustworthy and healthy data in the quantum IoT.

In order to enforce security in constrained environments, the HBS allows an \emph{adaptable selection of parameters} to enable trade-offs between signing speed and key size rather than using dedicated schemes. For instance, the key configuration involving underlying lightweight hash function and design optimization are suggested in~\cite{ghoshlightweight} for resource-constrained IoT. 

Through PRNG, HBS supports \emph{forward-secure construction} which implies that an attacker cannot subsume any information about previously used signature keys upon getting hold of the current private key. Forward-secrecy plays a consequential role in situations where devices can be tampered, compromised, or even stolen such as remote areas or outdoor device settings~\cite{palmieri2018hash}. 

\begin{table*}[ht!]
  \centering
\caption{Pros and Cons: Stateless vs. Stateful hash-based signature schemes.} \label{tab:proscons}
  \begin{tabular}{|p{1.2cm}|p{4.5cm}|p{4.5cm}|p{4.5cm}| }
  \hline
\textbf{Type} & \textbf{Pros} & \textbf{Cons} & \textbf{Use case}   \\ [0.5ex]
 \hline
\hline
\multirow{6}{*}{Stateful}     &
\begin{itemize}
    \item Shorter signature size
    \item Faster signature generation time
    \end{itemize}
& \begin{itemize}
    \item State synchronization problem (synchronization failure)
    \item Face cloning problem (volatile and non-volatile)
\end{itemize}
& Performance-constrained environment\\
\hline
\multirow{6}{*}{Stateless} &
\begin{itemize}
    \item No state synchronization problem
    \item No cloning problem
\end{itemize}
& \begin{itemize}
   \item Longer signature size
    \item Slower signature generation time
\end{itemize}
&Resource-constrained environment\\
 \hline
\end{tabular}
\end{table*}

\section{Introducing Hash-Based Signatures in the IoT Ecosystem} \label{HBSSIOT}
Improving data integrity of IoT devices against large-scale quantum computers stems from multiple factors, for instance, careful selection of HBS grounds for the underlying application requirements, device constraints, and design optimization criterion.
In this section, we highlight why quantum technologies matter in critical infrastructure and IoT. Furthermore, we discuss various factors while choosing the apt HBS from the get-go to avoid the digital transformation pitfalls of cutting-edge technologies.

\subsection{Stateless or Stateful?: Adoption of Apropos HBS Schemes} \label{aproposHBS}
The first factor is the adoption of apropos HBS for IoT devices. Before going into further details, the crux of stateless or stateful HBS is as under. The concept of statefulness arises from the use of one-time signature key pairs. As the robustness and security of HBS schemes depend completely on the use of non-repeated one-time key pairs; tracking the utilization of one-time key pairs is of paramount importance. To do so, one-time signing keys are used by following a sequential order such that an index or counter is stored in the global secret key to infer which one-time key pairs can still be utilized for signing purposes. In addition to the index, HBS schemes also include an authentication path that denotes a sequence of intermediate nodes required to reconstruct the path to the root node to validate a one-time public key against the global public key. In particular, different approaches consider different elements, for example, nodes for the next authentication path or pre-computed nodes as part of storing state data. 
For storing the state information, the size requirement depends on the tree structure, for instance, a 4-byte and 8-byte value is sufficient for XMSS and XMSS\textsuperscript{MT}, respectively. Thus, maintaining state information including the authentication path and the key index along with each signature equalizes the signing time. Nevertheless, it requires to store updated state information depending on the used parameters and implementation choices. 

On the other hand, stateless HBS schemes do not require to maintain the use of non-repeated key pairs; however, their signature sizes are significantly higher (as shown in Table \ref{tab:schemes}) making them impractical in some scenarios.
Thus, the optimal selection of a stateful or stateless scheme for embedded systems primarily depends on the \emph{time-memory} trade-off. For instance, stateful schemes exploit memory to store state information and have better run-time, hence, are well-tailored for performance-oriented systems while stateless schemes exploit processing power and have better memory utilization, hence, are well-suited for memory-constrained systems. 
It can be concluded that the stateful versions of HBS schemes offer better performance than the stateless versions, but require careful implementation to thwart an attacker to exploit the vulnerabilities related to state management. Summarized comparison of the pros and cons of both schemes is presented in Table~\ref{tab:proscons}.

 \begin{figure*}[!ht]
\centerline{\includegraphics[width=5.85in]{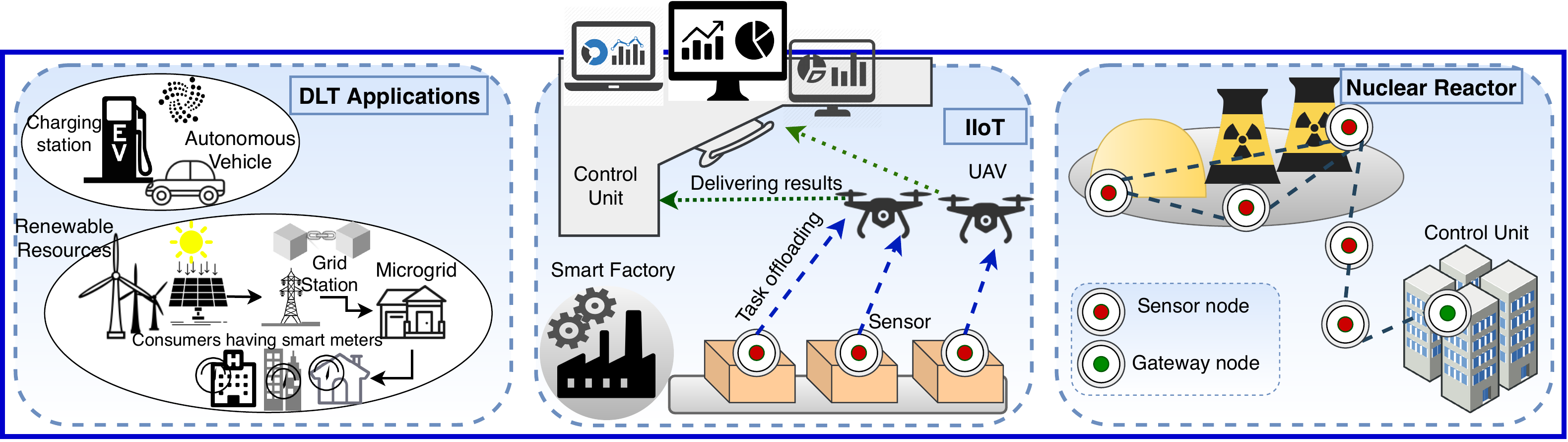}}
\caption{IoT use cases illustrating performance-constrained and resource-constrained scenarios.}
\label{fig:usecase}
\end{figure*}

\begin{figure}[!htpb]
\centerline{\includegraphics[width=2.85in]{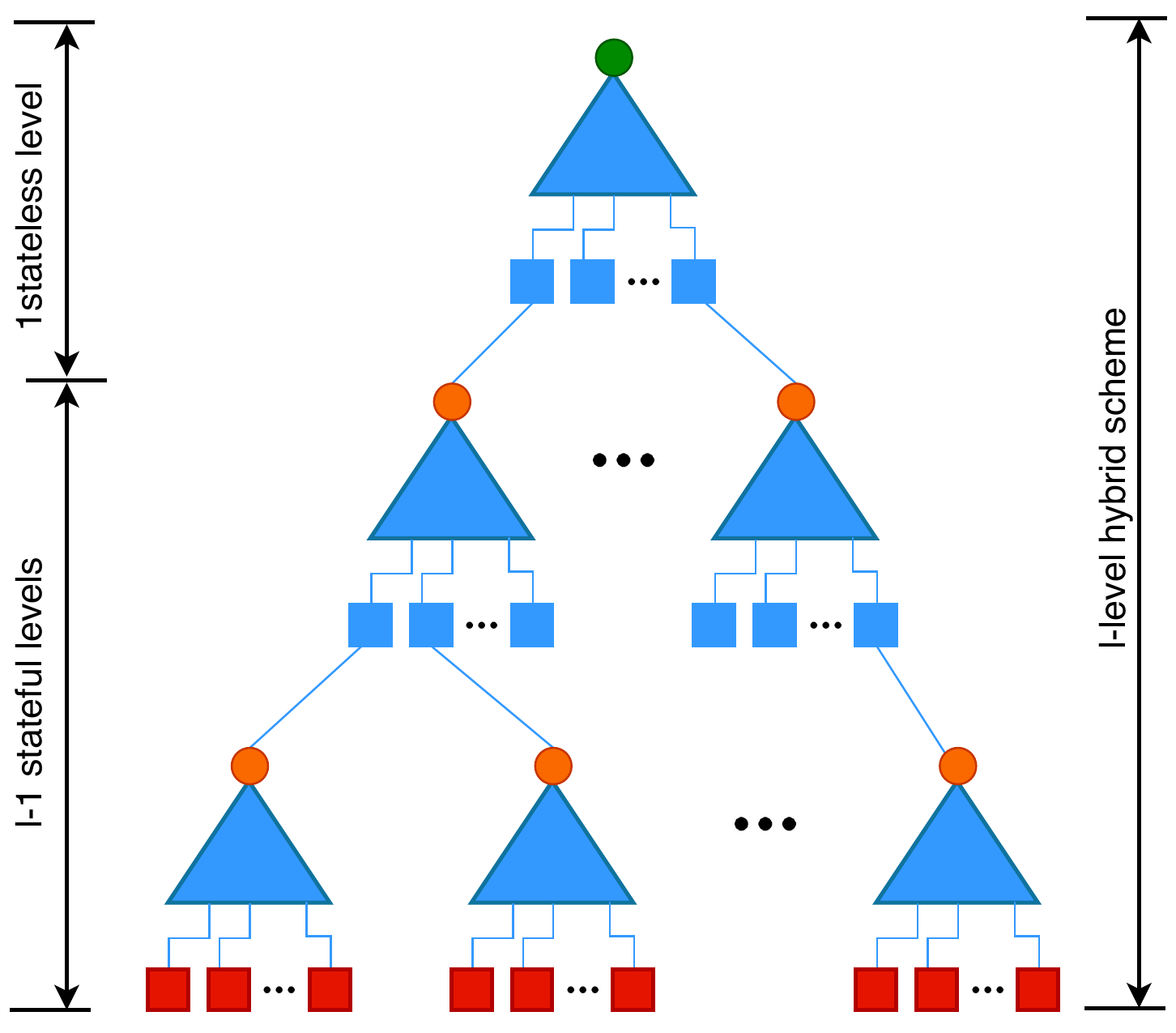}}
\caption{Combining a \emph{stateless} signature scheme (such as HORS-T) at the root level and a \emph{stateful} scheme (such as LMS or XMSS) at the lower levels: A \emph{hybrid} approach. (Figure source:~\cite{butin2017hash}).}
\label{fig:hybrid}
\end{figure}

For a given IoT system, the optimal selection of a stateless or stateful HBS scheme must be carefully weighted based on the fact that whether the system is \emph{performance-constrained} (processing time, computational complexity) or \emph{resource-constrained} (energy usage, memory consumption). 
For instance, consider a nuclear reactor (as shown in Fig.~\ref{fig:usecase} (right-most)) where sensors (for instance, temperature, flow, pressure or level) are deployed in order to monitor (heating system, water pressure, or water level). The sensors' readings are notified to a control room that is accountable for making critical decisions (to turn on/off any valve or to adjust any values) based on the sensors readings. Under such a performance-constrained environment, the integrity and authenticity of data must be verified efficiently because operations such as parameter tuning, data debugging, and aging management rely on data-driven decisions.  
Similarly, Fig.~\ref{fig:usecase} (middle) illustrates an example of Industry 4.0 that exhibits a synergy between industry and IoT. The example shows a smart factory where fog-enabled Unmanned Aerial Vehicles (UAVs) can be used to gather the tasks from the sensor, compute the tasks, and deliver the processed results to the control unit.
Under the resource-constrained Industrial Internet of Things (IIoT), the tasks are offloaded to UAVs to conserve sensor energy. Following a better approach, stateful schemes are suitable candidates for the former case while stateless schemes are apt for the latter case. 

A setting under which both resource-constrained and performance-critical IoT systems are desirable, a reasonable compromise between stateful and stateless schemes is the hybrid approach. For instance, in \cite{mcgrew2016state}, the authors proposed a hybrid method by combining the stateless signature scheme such as HORS-T and the stateful signature schemes (e.g., XMSS and LMS) at the root level and the lower levels, respectively (as shown in Fig.~\ref{fig:hybrid}). 
Such strategy overcomes downsides while merging the benefits of both stateful and stateless HBS schemes.

\subsection{Implications of Quantum Computing on DLT}
The second factor covers the imminent prodigious threats to the applications of DLT by quantum computers.
DLT despite being a quintessence solution of the Internet of Everything (IoE), one of the main challenges is the reliability of the data generated by \textit{things}. DLT can ensure the immutability of data in the ledger, nevertheless when the data generated by IoT devices is dubious or malicious due to the physical environment, participants, vandalism, and the failure of the devices, then its further propagation through the ledger stays corrupted. Furthermore, the analysis and interpretation based on such abnormal data produce catastrophic results, especially for applications relying on data for critical decision-making processes, risk assessment, and performance evaluation~\cite{suhail2020provenance}. The corrupted devices either face physical damage or limit the firmware updates to refrain them from actuating over possible bugs or security breaches. 
One such solution to ensure the trustworthiness of data by the device in question is to keep track of data lineage through data provenance~\cite{suhail2020provenance}.

In the IoT ecosystem, blockchain is another ahead of the curve DLT solution that has powered resource-consuming devices to participate in Machine-to-Machine (M2M) or Machine-to-Human (M2H) economy autonomously, for instance, to support and accelerate the distributed energy in a microgrid or electric vehicle charging (as shown in (Fig.~\ref{fig:usecase} (left-most)). Currently, most of the blockchain-based solutions heavily rely on conventional cryptographic standards to support the immutability and transparency of data.
However, ledgers that are not quantum-resistant could pose long-tail data risk. High-powered quantum computers can jeopardize M2M or M2H world by potentially enabling attackers with quantum computers to monopolize the network by sabotaging transactions and preventing their own transactions from being recorded or double-spend~\cite{fedorov2018quantum}. To prepare for the quantum apocalypse, blockchain-enabled schemes that already support post-quantum techniques are Quantum Resistant Ledger (QRL)~\cite{QRL} (using XMSS), IOTA \cite{popov2017iota} (using WOTS), and Corda (using BPQS: a single-chain variant of XMSS). 

\subsection{Optimal Design Objectives}  \label{design_obj}
The third factor is the optimized design objectives for IoT devices. In particular, function independence characteristic of HBS schemes make them a suitable candidate for ultra-constrained IoT settings, for instance, the \emph{latency-area} optimized design approach proposed in~\cite{ghoshlightweight}. Similarly, other design trade-offs for IoT devices include a \emph{lightweight hash function} for energy-efficient computation of signature/verification operations. For instance, in~\cite{ghoshlightweight}, the authors implement and perform explicit area and latency analysis of four hash candidates including SHAKE-256, SHA-256, S-quark, and Keccak-400. Considering energy budget constraints, Keccak-400 is selected. 

To meet the design objectives of resource-constrained IoT nodes, in addition to smaller parameters and light-weight hash function, appropriate algorithms based on the design specification of motes are needed. Such co-design principles based on hardware and software provide a trade-off between area overhead and hardware penalizing. For example,~\cite{ghoshlightweight} proposed a scheme in which WOTS\textsuperscript{+} operations are defined at the hardware level due to a significant amount of repetitive hash computations and to yield smaller footprints while XMSS operations and WOTS\textsuperscript{+} parametrization control are defined at the software level to preserve latency gains.

Another design aspect essential to all HBS is the \emph{generation of either hardware-based or software-based random numbers}. 
Keeping in view that the sources of external entropy are limited for critical IoT deployments in the isolated environment, hardware-based random numbers are preferred (for instance, \emph{Quantum Random Number Generation} (QRNG) \cite{Collantes2017} chip). QRNG is a physically and provably secure source of randomness in contrast to PRNG that requires monitoring to maintain sufficient randomness for business protection as adversaries commit additional resources to find patterns in PRNG implementations.

\subsection{Potential Attacks on HBS}
The fourth factor is handling of the attack surface even in the presence of quantum-resilient signature schemes, for example, evaluating the HBS in the presence of physical (or implementation) attacks, i.e., side-channel attacks and fault attacks.
In a differential side-channel attack, the attacker gains extra information by eavesdropping on a side channel, for instance, power-monitoring, electromagnetic leaks, or processing timing during the computation of the signature. 
Whereas in a fault attack, a fault, which can be either natural or malicious, is misbehavior of a device that causes the computation to deviate from its specification. The goal of the attacker is to exploit such information to gain access to the secret. 
HBS schemes are vulnerable to hardware fault attacks both in the presence of natural and malicious faults. To address fault-attack resistance, in \cite{mozaffari2017fault}, the authors present an implementation approach to make stateless hash-based constructions more reliable against natural faults and malicious faults. 
Similarly, in \cite{kannwischer2017physical}, the authors discuss implementation recommendations for XMSS to resist implementation attacks (for example, selection of side-channel resistant PRNG, computation of optimized authentication path, and strategy for caching of signatures). In addition, the proposed scheme can be tailored based on the reliability objectives and available resources~\cite{cheng2017securing}.

\subsection{Benchmark: Software and Hardware}
The fifth factor is the benchmark for evaluating the performance of HBS. From the software benchmark perspective, the run-time of key generation, signing, and verification processes whereas from the hardware perspective, CPU cycles, key size, signature size, and energy consumption are among the targeted evaluation metrics. In general, the parameter sets are highly dependent on the underlying construction of a particular scheme. 
For software benchmarking, frameworks such as System for Unified Performance Evaluation Related to Cryptographic Operations and Primitives (SUPERCOP) and ECRYPT Benchmarking of Cryptographic Systems (EBACS) are commonly used for the evaluation of the software performance. For hardware benchmarking, Application-Specific Integrated Circuit (ASIC), 
Field-Programmable Gate Arrays (FPGA), or other micro-architectures can be configured and programmed accordingly. Also, architecture-specific optimizations such as Advanced Encryption Standard New Instructions (AESNI) or Advanced Vector Extensions 2 (AVX2) instructions are used to make it implementable on the available micro-architecture~\cite{bernstein2019sphincs+}.

\subsection{Trust Chain: Combining HBS and Provenance}
The sixth factor involves the combination of HBS schemes and data provenance that epitomizes the importance of \emph{trustworthy data}.
On one hand, HBS ensures the accuracy, fidelity, availability, and confidence of data, whereas on the other hand data provenance identifies the sources behind stale, latent, and tardy data. Therefore, such combination can solve the problems related to erroneous or faulty data thereby enhancing the quality of data. Another instructive use case of such a scenario is the supply chain where data integrity and provenance supplement each other to solve the traceability problems, counterfeit concerns, and data accessibility issues in the supply chain space~\cite{suhail2019orchestrating}.

\subsection{Establishing End-to-End Security}
The seventh factor is to establish horizontal end-to-end security.
A reliable infrastructure is a must to boost the end-to-end ecosystem's security especially in the presence of a diverse range of cybersecurity threats (such as data breaches, (D)DoS attacks, and so on) and continuously increasing demands of efficient communication requirements (such as ultra-reliability, low-latency). Though 5G promises to solve most of the communication requirements for many versatile applications, for example, tactile Internet, massive IoT, autonomous vehicles, and many more. However, the inherent security flaws still need more attention, for instance, location tracking, activity profiling, etc. Similarly, some other quantum-linked features such as quantum-safe communication, quantum Internet, and \emph{Quantum Key Distribution} (QKD) also require a due attention to deftly integrating quantum computing in the fabric of 5G and beyond.

Hence, application-specific and platform-dependent trade-offs must be considered with regards to signing speed, signature size, a desired number of signatures, memory constraints, processing limits, light-weight underlying hash functions, and hardware support for particular hash functions.

\subsection{Current Industry-scale Implementation Efforts}
Albeit with a restricted number of qubits, quantum computers already exist though luckily for today’s security cannot run Shor’s algorithm. For example, a Canadian company, \emph{D-Wave Systems} was the earliest to market and has already launched its 2000Q System quantum computer~\cite{Dwave}. \emph{IBM Q Quantum Computation Center} is an industry-first initiative to build commercial universal quantum systems for business and science applications~\cite{IBMQ}. Furthermore, \emph{Google} claimed to have achieved the \emph{quantum supremacy} by introducing a superconducting quantum processor called \emph{Sycamore}~\cite{arute2019quantum}. According to their benchmark task, Sycamore outperforms (took 200 seconds) state-of-the-art supercomputers that would require approximately 10,000 years to perform a random sampling task. To continue the benchmark progress, IBM upends Google's claim and experimentally proved that the same task can be performed on a classical system in 2.5 days by incorporating other conventional optimization techniques to improve performance~\cite{pednault2019leveraging}. 
In addition to these, other companies participating in the race of developing quantum computers include Intel, Microsoft, IonQ, to name a few. 
Such back-to-back research efforts by tech-giants herald a degree of technical maturity towards a quantum leap which ultimately opens new frontiers for quantum computing in the IoT world.

\section{Standardization Efforts and Future Research Challenges of HBS Schemes in the IoT} \label{challenges}
In this section, we highlight the standardization efforts carried out for HBS schemes and future research challenges.

\subsection{Standardization Efforts} 
\label{standard}
The efforts to solicit and evaluate quantum-resistant public-key cryptographic algorithms for an inevitable transition to post-quantum cryptography are underway by many standardization organizations. 
For instance, the National Security Agency (NSA) plans to shift from the Suite B set of cryptographic algorithms towards post-quantum cryptography~\cite{NSA2015}. 
Furthermore, workshops and calls for proposals are initiated by the US National Institute of Standards and Technology (NIST)~\cite{chen2016report} in the Post-Quantum Cryptography Standardization project (evaluation of Round 2 candidate algorithms in the process~\cite{nistList}) and European Telecommunications Standards Institute (ETSI)~\cite{pecen2014quantum} in Quantum-Safe Cryptography (QSC)~\cite{QSC} project to indicate the increasing necessity of switching to post-quantum cryptography. Regarding the specification of HBS, Internet Engineering Task Force (IETF) is targeting both XMSS and LMS for standardization~\cite{mcgrew2017internet, hulsing2017internet}. 
Other ongoing projects and developments to promote research on post-quantum cryptosystems by European Commission include PQCRYPTO~\cite{PQcrypto} (conducting research on post-quantum cryptography for small devices, the Internet, and the cloud), SAFEcrypto~\cite{safecrypto} (focuses on secure post-quantum cryptographic solutions to preserve the privacy of government data, and protection of data in communication systems)~\cite{cheng2017securing}. Similarly, the CryptoMathCREST~\cite{CryptoMathCREST} research project is supported by the Japan Science and Technology Agency to study the
mathematical problems underlying the security of PQC. 

\subsection{Future Research Challenges} 
\begin{figure*}[ht!]
\centerline{\includegraphics[width=4.0in]{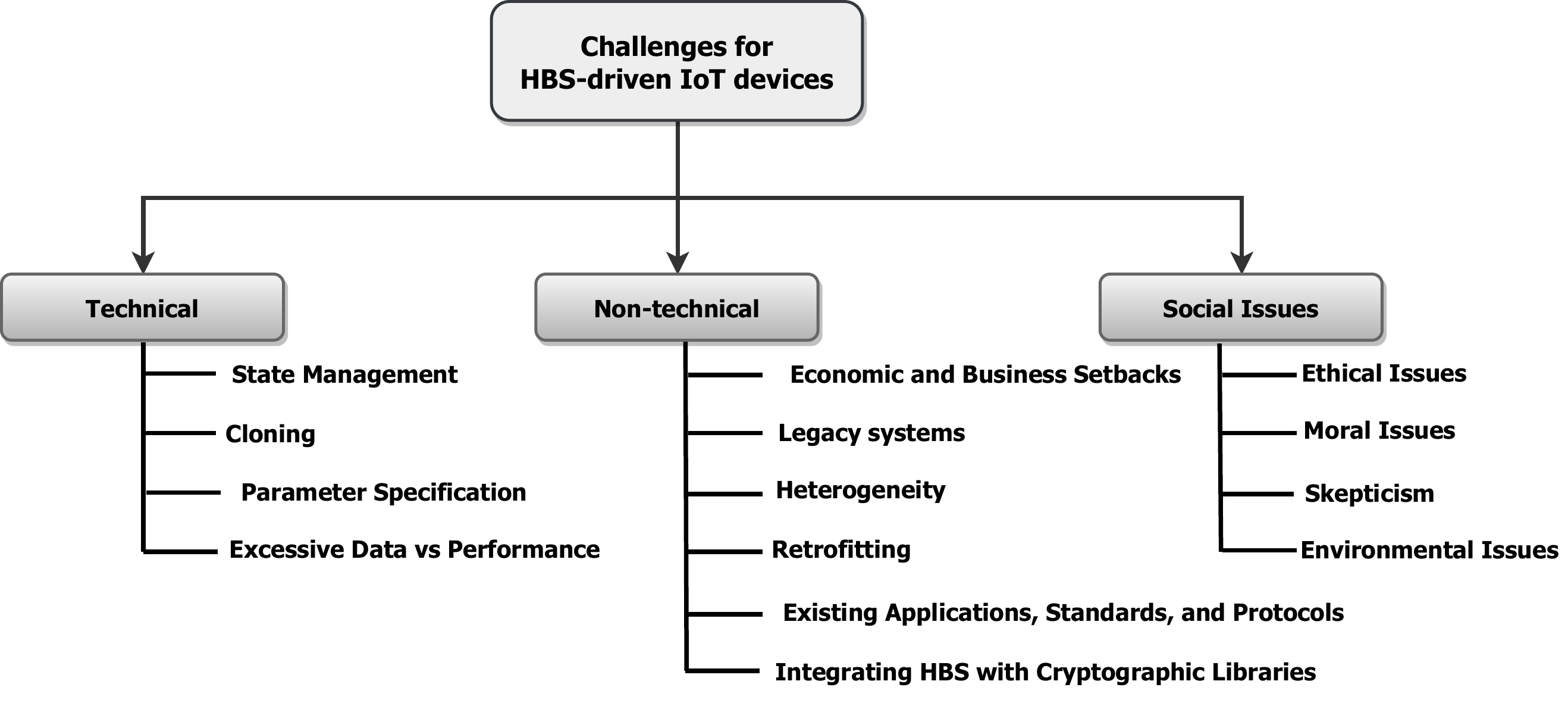}}
\caption{Current and future challenges for HBS schemes in the IoT domain.}
\label{fig:challenges}
\end{figure*}

In the quest to secure IoT in the quantum era, following technical, non-technical, and social challenges of HBS schemes call for further investigation. We also outline the key recommendations necessary to act and prepare for the quantum era. Fig.~\ref{fig:challenges} presents the detailed taxonomy of the current and future research and deployment challenges for HBS-driven IoT and we summarize the
challenges along with causes and possible solutions in Table~\ref{tab:technical_non-technical}. 

\subsubsection{Technical Challenges} 
\label{technical}
Here we discuss technical challenges related to IoT devices with reference to quantum computing.

\emph{a) State Management}:
In the stateful signing algorithms schemes, state management is one of the challenging snags to the widespread use of HBS schemes. In this problem, the version of the private key in non-volatile memory (disk) must be continuously synchronized with that in volatile memory (RAM) to avoid key synchronization failure. Crash of an application or an operating system, corruption of the nonvolatile state, power outage, or a software bug could be among the potential causes of synchronization failure~\cite{mcgrew2016state}. The delay caused by the synchronization of the private key between the storage unit and execution unit results in additional latency for the signature generation time, thus highly deteriorating the overall performance of the system.

\emph{b) Cloning}: 
Another problem in the stateful signature scheme is cloning. Such type of risk occurs when a private key is copied and then used without coordination with execution units (known as non-volatile cloning) or without coordination with storage units (known as volatile cloning).
Live Virtual Machine (VM) cloning or restoration of a key file to a previous state from a backup system could potentially cause volatile or non-volatile cloning. The cloning problem results in the generation of multiple signatures from the same system state, thus crucially undermining security. For instance, in case of live VM cloning, values that may only be used once, are at risk, including initialization vectors, pseudorandom numbers, counters for encryption, one-time passwords, and seeds for digital signatures~\cite{everspaugh2013virtual}. Similarly, initial sequence numbers could be reused for hijacking in the case of the S/Key (a one-time password system)
and the TCP protocol~\cite{mcgrew2016state}. 
Issues with such primitives can be problematic even for classical digital signature schemes. To summarize, nonvolatile cloning may not cause any issue to a system devoted only to the signature generation; however, it can cause significant risk to the general-purpose software system. On the other hand, volatile cloning leads to catastrophic results particularly due to the vulnerabilities pertaining to caching of random numbers. 
Therefore, tailored to specific use-cases, the state management strategies must be gauged in a nuanced way. 
For instance, resource-constrained sensor nodes piggyback on UAVs for computation and processing of tasks (as shown in Fig.~\ref{fig:usecase} (middle)). In this scenario, the issues of state management (either key synchronization or cloning risk), may cause problems including (i) performance issues at delivering results to control unit, (ii) energy issues at UAVs, and (iii) data integrity risks at the control unit.

Though stateless signing algorithms solve the state and key synchronization concerns; however, signature size is still a problem. To resolve the issues of stateless and stateful schemes, a hybrid approach discerns the essential worth with smaller signatures and faster signing deserves further exploration. Other possible solutions suggested in~\cite{mcgrew2016state} include state reservation strategy and hierarchical signature schemes. Simply put, in a state reservation approach, the private key that is ahead of the current signature among the available $N$ signatures is written back into storage, thereby avoiding the need to write the updated private key into nonvolatile storage. In the case of a hierarchical signature scheme, a volatile bottom level enforces the reservation property such that the private key of the volatile level is not synchronized in nonvolatile storage. Such combined volatile/nonvolatile hierarchical signature scheme property avoids synchronization problems and is considered a reasonable model for problems related to writing operation scenarios such as power outage or crash of an application. However, both of these solutions do not address the nonvolatile cloning problem.

\emph{c) Specification of Parameters}:
Another issue is that the universal specification of a parameter set highly depending on the intended use-cases. Since constraints on performance aspects such as signing speed and key size are highly dependent on underlying use-cases, therefore, it is hard to define one universal parameter set for every scenario. For example, software update authentication does not entail high-frequency signing, however, the converse is true for Hypertext Transfer Protocol (HTTP) over Transport Layer Security (TLS). Another example is the individual user's email signing that does not require frequent signing though, however, keeping in view the usability considerations, the priority is given to the signature size to limit message expansion~\cite{butin2017hash}.
 
HBS schemes need to offer concrete parameter choices to provide user guidance while considering constraints on performance aspects such as signing speed and key size. For concrete instantiations, proper guidance (rules and regulations with concrete steps) should be included in standards. In this regard, \cite{hulsing2015xmss, mcgrew2017internet} suggest concrete parameter sets and discuss the crucial
element of security levels for the proposed parameter sets, however, unable to address their adequacy tailored to specific applications. Thus, the use of underlying hash function, state management strategies and other construction parameters must be evaluated in a nuanced way depending on the intended use-case as also discussed in subsection~\ref{aproposHBS} and~\ref{design_obj}. Though the recommended parameters should be provided by the cryptographic community; however, customizing signing speed, key size, and other construction parameters depending on the application scenario is a crucial asset.

\emph{d) Trade-off Between Excessive Data and Performance}:
Depending on the application and underlying infrastructure, a network of things may have a dynamic and rapidly changing dataflow and workflow where data inputs are provided from a variety of sources such as sensors, external databases or clouds, and other external subsystems. As the generation of vast amounts of data over time renders IoT systems as potential \emph{big data generators}, in this regard, \emph{how can we ensure the speed and performance of underlying HBS schemes?} One potential solution is to adopt hybrid HBS schemes to enable a trade-off between performance-constrained and resource-constrained environment. Besides, more efficient algorithms may open the way to application in the diverse and constrained reality of the vast majority of IoT devices.

\subsubsection{Non-Technical Challenges}
Bringing quantum computing could enable advances in many futuristic technologies; however, it requires consideration of many significant factors. Here, we discuss non-technical challenges related to IoT devices with reference to quantum computing.

\emph{a) Business and Economic Setbacks}:
As the age of quantum computing is gradually dawning, it seems that new hardware systems are among constantly increasing requirements. Therefore, the question, \emph{how IoT devices can adapt to quantum computing with their current embedded hardware (such as crypto-processors) that is generally optimized to carry out certain cryptographic operations?}, must be answered. However, the implications of such demands are likely to face huge business and economic setbacks in terms of expenditures on new or upgraded IoT infrastructures to handle the increased workload. Thus, bringing in new hardware may be too expensive for cost-sensitive large-scale applications that are usually looking forward to cost-effective solutions by drastically reducing capital expenditure (CapEX) and operational expenditure (OpEX).

\emph{b) Entanglement in Legacy Systems, Existing Applications, Standards, and Protocols}:
In addition to the aforementioned demand for new hardware systems, one of the substantial concern is \emph{how to retrofit legacy systems with advanced security solutions?} Because shifting to novel quantum-based infrastructure for IoT demands fragile engineering environment, for example, temperature constraints for operating quantum infrastructure, the limited range for terrestrial quantum communication networks, the staggering cost of various hardware for carrying out QKD, budget funding, and other obstacles that may limit the usability of quantum-based systems at the moment. 

Similarly, another question is \emph{how existing applications and protocols can adapt to quantum computing with their current standards?} One solution is to modify the existing protocols to handle larger signature or key size by segmenting the data into multiple massages for bandwidth-constrained applications (e.g., self-driving cars). However, the status quo will change as new applications and protocols must set their standards keeping in mind the demands of quantum schemes.
Existing protocols might need to be modified to handle larger signatures or key size, for example, through the segmentation of messages. Also, protocol designers should be aware that changes in the underlying cryptography may certainly be necessary for the future, either due to quantum computing or other unforeseen advances in cryptanalysis. For new applications, implementations must keep the demands of PQC in mind and allow the new schemes to adapt to them as PQC requirements might shape future application standards.

\emph{c) Heterogeneity in Terms of Application and System}:
Another unique characteristic of IoT devices is heterogeneity. On one hand, heterogeneity may appear in terms of divergent application requirements, for instance, resource constraints in sensor networks, security constraints for medical implantable devices, performance constraints for IIoT, etc. On the other hand, it may appear in terms of diversified architecture requirements, for instance, interoperability across diverse platforms from different vendors, integration of disparate sub-systems, and the existence of compatibility among sub-systems to work in conjunction without conflict. Possible solutions to handle heterogeneity is to consider interoperability and integration of systems or subsystems and to promote flexibility and include abstractions to facilitate integration among existing applications
and libraries. The systems that have prescriptive requirements such as military-critical and safety-critical systems must consider all of these aspects while enforcing appropriate quantum-resistant algorithms upon careful identification of the system requirements (such as performance contracting).

\emph{d) Bridging The Gap: Integrating HBS with Well-known and Tested Cryptographic Libraries}:
Integrating HBS with well-known and tested cryptographic libraries plays an ergonomic role to ensure the wide availability of HBS in security infrastructures and serves the goal of absolute security shared by all stakeholders. Though in the case of HBS, proof-of-concept implementations exist such as~\cite{buchmann2011xmss, hulsing2012forward} which mark a necessary step towards their widespread usage. On a related note, such stand-alone implementations are unable to facilitate both technical interfacing and strategic decisions such as parameter selection. In \cite{butin2015real}, the authors suggested avoiding case-by-case implementation of cryptographic primitives as it is inopportune for organizations to develop their own specific ad hoc implementations and recommended the usage of commonly used software cryptographic libraries (such as Open SSL) particularly because of their ability to include abstractions to facilitate system integration and combination.

\subsubsection{Social Challenges}
In the following, we discuss the social challenges faced by HBS in IoT networks.

\emph{a) Ethical and Moral Consequences}:
The access to large-scale quantum computers by the government institutions and other research funding organizations can be analyzed from ethical perspectives. For example, if access to quantum computers is limited to a few government agencies, they may dominate or dictate other nations (also referred to as the \emph{Big Brother Problem}). Also, considering the risk that only a few big companies or corporate laboratories are able to afford quantum computers due to massive investment, the entrenched giant companies may use the efficiency gains to out-compete their competitors and thus lead to monopolies or oligopolies~\cite{de2017potential}. Even worse, the enterprises may use it with criminal intent such as industrial espionage for competitive advantage, mass-surveillance, and other undesirables. Furthermore, evildoers can harvest high-value data (such as medical data or sensitive government data) now and break it later by using quantum computers. 
The best way to make the impact of quantum computers positive is to enable their wide accessibility to people to run programs on them through the cloud. A toy version of such an idea with a 5-qubit computer through the cloud is provided by IBM’s \emph{Quantum Experience}~\cite{IBMquantum}. Similarly, to access quantum computing ecosystem platforms should be provided to enable academic researchers who are focused on theoretical work and tech-industry experts who are familiar with real-world performance needs and security demands to collaborate and share their experiences.

\emph{b) Skepticism in Quantum Computing}:
On one hand, there is an on-going race to build universal quantum computers along with a huge amount of scholarly literature and awareness about the potential societal impact on the breaking down of current-grade cryptography. On the other hand, the physical realization of quantum computers has been a hard slog that eventually raises serious doubts by quantum skeptics. The skeptics argued the possibility to build a scalable quantum computer due to various factors (such as noise, constraints on state preparation, unreliability, virtuous cycle, manufacturing errors, etc.) though they do agree that theoretically quantum computation does offer an exponential advantage of classical computation~\cite{vardi2019quantum}. Gil Kalai, one of the most prominent quantum skeptics also argue against quantum computers due to several underlying facts related to noise in physical systems and quantum error correction~\cite{Gil}. 

According to the analysis given by~\cite{national2019quantum}, quantum computing needs to create a virtuous cycle, similar to that of the semiconductor industry, in order to generate a commercial demand by attaining sufficient economic impact and to fund the development of increasingly useful quantum computers as a major milestone. The same quandary goes for IoT devices, for instance, how ultra resource-constrained devices are going to adopt compute-intensive schemes, how to upgrade or replace IoT devices to carry out quantum-secure algorithms, etc. Also, from the software perspective, software developers must have enough knowledge of quantum theory to write code for the machines as quantum algorithms require a completely different way of thinking about problem-solving.
In a net shell, keeping in view a rudimentary stage of evolution of quantum computers (in terms of hardware and software), most of the scientists are of the view to wait and see as a lot of work is needed to build post-quantum systems that are widely deployable while at the same time inspiring confidence.

\emph{c) Environmental Aspects}:
The computational and processing time required by the signing algorithm highly impacts the energy consumption by resource-constrained IoT devices which could ultimately make a somewhat noticeable environmental impact as the number of devices connected to the Internet is exponentially growing. 
To curtail such an impact on the environment, efficient signature algorithms should be used so to conserve energy which is beneficial for both scientific interests and environment interests~\cite{sjoberg2017post}.
Another environmental aspect is the upsurge in e-waste caused due to new hardware (such as crypto processors) as the existing devices or embedded components may not be able to efficiently go hand in hand with the quantum-safe algorithms. Moving to quantum-resistant crypto primitives which involve more computationally-intensive tasks may affect the performance of the current systems and even render some devices or components obsolete.

\subsubsection{Thinking Ahead: A Pragmatic Approach}
While we are still preparing for quantum-safe algorithms, but at the same moment, we have to protect the information that is already vulnerable; therefore, the overarching question is that, which defensive strategies should be adopted by the government to avoid significant geopolitical and diplomatic ramifications and corporate organizations to mitigate potential liabilities? In the following, we outline a few prudent measures and laying groundwork that must be adopted by the organizations to plan and prepare a quantum-secure IoT infrastructure.

\begin{itemize}
    \item Firstly, identify and document information assets (including business value, access control, data sharing arrangement, handling at end-of-life, backup and recovery procedures) and the current cryptographic protections (such as lengthening or maximizing
current public key sizes) to determine the organization’s vulnerability to external and internal threats. Then the next step is to document the threat models and threat actors as follows:
    \begin{itemize}
        \item The threat models encompass critical infrastructure deployments and high inter-connectivity and inter-dependencies among devices, subsystems, and external third-party systems. The models must also recognize the requirement of lifetime systems that stretch over decades while others may refresh annually or more frequently. 
        \item Identify threat actors and estimate their timeline to access and exploit quantum technology.
    \end{itemize}
    
    \item Secondly, a continuous evaluation based on an estimation of the lifecycle and field deployment conditions for such threat models is required as new technologies and attack vectors emerge.
    \item Thirdly, investigate the impact of quantum technologies and conduct a Quantum Risk Assessment (QRA) on the underlying systems. In this regard, any cyber risk assessment must be periodically updated to account for emerging threats and to take advantage of improved security solutions as quantum technologies are not mature yet and are still rapidly evolving. 
    \item Fourthly, build crypto agility into systems to ensure an upgrade path and the ability to conduct remote upgrades in a secure, timely and pro-active manner. 
    \item Fifthly, 
    \begin{itemize}
        \item from the hardware perspective, build devices and systems with long term security in mind, for instance, hardware-based key generation for adequate security of cryptographic operations throughout the lifetime of the device in the field. Another long-term solution could be to rely on quantum cryptographic methods to reduce hypothetical risk to business processes until quantum computing hardware becomes commoditized into solutions. 
        \item from the software perspective, if possible, finding other PQC algorithms that can be used as drop-in replacements to make the transition less disruptive, 
       
        \item software-as-a-service or third-party platform providers can also be commissioned for further assistance,
       
        \item perform the cost estimation of new or upgraded hardware and software systems. This may also involve equipping the organization personnel with practical quantum skills or even accessing a platform to learn world-class expertise and technology to advance the field of quantum computing.
    \end{itemize}
       \item Finally, after identifying and prioritizing the activities required to shift the organization's technology to a quantum-safe state, keep track of governance infrastructure and migration plans that are required to respond to changes into systems in order to address immediate concerns while permitting the federation of new quantum technologies.  
\end{itemize}
  
Thus, now and in the future, strategic thinking and long-term planning in terms of short-term remedies and small-scale fixes to repercussions of vulnerable information must be adopted for protecting sensitive information at banks and government databases until quantum-safe schemes will become fully available with pragmatic solutions and current infrastructures are rendered void.

\section{Conclusion} 
\label{conclusion}
The countdown of the nascent quantum computing paradigm commenced upon the realization of security threats to classical digital signatures schemes. This hype cycle also surges in the IoT world in order to draw attention to the security, authenticity, and integrity of sensory data. To address such issues, HBS is considered to be part of the future portfolio of deployed PQS particularly due to their minimality of the required security assumptions.

In this article, we covered different aspects of HBS schemes including their classification, along with their underlying construction parameters, and striking features. We focused on the problem of introducing HBS schemes in the IoT ecosystem, wherein we highlighted the adoption of suitable schemes considering application-specific (such as signature size, signing speed) and platform-dependent (such as memory constraints, hardware support for specific hash functions) trade-offs. Furthermore, we also identified a set of future research challenges with an open-ended discussion in the adoption of HBS schemes by the IoT community.  We hope that this survey provides close insights to researchers to overcome the challenges and pave the way for the standardization of HBS schemes in IoT-based applications.

As a part of our future work, we plan to investigate other post-quantum signature schemes, compare and evaluate them in terms of various construction parameters that are necessary for secure, resource-constrained, and performance-constrained IoT environment.

\clearpage
\onecolumn

\begin{longtable}{|p{4.0cm}|p{5.5cm}|p{5.5cm}|}
\caption{Current and future research and deployment challenges in HBS-driven IoT.} \label{tab:technical_non-technical} \\
  \hline
  
\textbf{Class} & \textbf{Key challenges}  & \textbf{Possible solutions }      \\
    \hline
  
\endfirsthead
\caption{Current and future research and deployment challenges in HBS-driven IoT.} 
 \\
  \hline
\textbf{Class} & \textbf{Key challenges}  & \textbf{Possible solutions }     \\
\hline
\endhead
\multicolumn{3}{r}{\footnotesize\emph{continued on the next page}}
\endfoot
\endlastfoot

 \multicolumn{3}{|c|}{\textbf{T: Technical challenges  \hspace{10pt}  NT: Non-technical challenges \hspace{10pt} S: Social challenges}} \\ 
 \hline
 \hline

\textbf{T1:} State management & \begin{itemize}

\item Synchronization failure of the private key between non-volatile and volatile memory. 
\item Effecting the performance of the system, i.e., additional latency for the signature generation time.
\end{itemize}
& Use stateless or hybrid HBS schemes to avoid key management issues.
\\
\hline
\textbf{T2:} Cloning & Using a copied private key without coordination with execution units or storage units.  & Use stateless or hybrid schemes.\\
\hline
\textbf{T3:} Specification of parameters & Require use-case specific parameter set.  & Define standards for parameter set guidance for use cases. 
\\
\hline
\textbf{T4:} Trade-off between excessive data and performance & Dynamic dataflow in particular IoT applications.  & Use hybrid HBS schemes.\\
\hline
\textbf{NT1:} Business and economic setbacks & \begin{itemize}
  
\item How the current embedded hardware can adapt to quantum-safe cryptographic operations?
\item Upgrading or establishing new IoT infrastructures incurred a huge economic burden.
\end{itemize} & 
Need to identify and plan expenditure on software and hardware costs.
\\
\hline
\textbf{NT2:} Entanglement in legacy systems, existing applications, standards, and protocols & 
\begin{itemize}
\item How to retrofit legacy systems with advanced security solutions?
\item How existing applications and protocols can adapt to quantum computing with their current standards?
\end{itemize}& 
\begin{itemize}
\item Modify the existing protocols to handle larger signature or key size.
\item New applications and protocols must set their standards based on the demands of quantum schemes.
\end{itemize} \\
\hline
\textbf{NT3:} Heterogeneity in terms of application and system & How to provide a quintessential infrastructure for divergent application requirements tailored to specific use cases and diversified architecture requirements strictly depending on platforms and vendors. & \begin{itemize}
    \item Consider interoperability and integration of systems. 
    \item Adopt appropriate algorithm after carefully identifying the system requirements.
    \item Promote flexibility and include abstractions to facilitate integration among existing applications and libraries. 
     \end{itemize}
    \\
\hline
\textbf{NT4:} Integrating HBS with well-known and tested cryptographic libraries & \begin{itemize}
    \item How to ensure the wide availability of HBS in security infrastructures? 
    \item How to avoid case-by-case implementation of cryptographic primitives? 
    \end{itemize}
    & Promote integration of HBS with well-tested and commonly used cryptographic libraries.\\
\hline
\textbf{S1:} Ethical and moral issues & \begin{itemize}
    \item Government agencies having access to quantum computers may attempt to establish dominion over other nations.
    \item Colossal firms having quantum computers may monopolize the global market.
    \item Researchers and scientists may patent or even hoard knowledge, resulting in limited access to quantum computing knowledge.
    \end{itemize}  & 
    Encouraging widespread knowledge of the quantum computing paradigm in both academia and industry through collaboration.
     \\
\hline
\textbf{S2:} Skepticism &  \begin{itemize}
    \item Quantum skeptics doubts over the possibility to build a quantum computer due to noise in addition to other factors.
    \item How to generate a commercial demand of quantum computers? 
    \end{itemize} 
    &  
    \begin{itemize}
    \item Leverage standardized post-quantum cryptographic solutions to remain on safer side.
    \item Needs to create a virtuous cycle.  
    \end{itemize} 
    \\
\hline
\textbf{S3:} Environmental issues & \begin{itemize}
    \item Energy consumption by massively deployed IoT devices.
\item E-waste caused due to new hardware.
\end{itemize}  & \begin{itemize}
    \item Use of efficient algorithms to conserve energy.
    \item Retrofitting.
    \end{itemize}
    \\
\hline

\end{longtable}
\clearpage
\twocolumn

\ifCLASSOPTIONcaptionsoff
  \newpage
\fi

\end{document}